\documentclass[prl,showpacs,preprintnumbers,amsmath,amssymb]{revtex4}
\usepackage{graphicx}
\usepackage{bm}

\begin{document}

\title{Critical Fluctuations in the Microwave Complex Conductivity
of BSCCO and YBCO Thin Films}
\author{
D.-N.~Peligrad\thanks{present address: Philips Research
Laboratories, Weisshausstrasse 2, D-52066, Aachen, Germany},
B.~Nebendahl\thanks{present address: Agilent Technologies
Deutschland GmbH, Herrenberger Str. 130, D-71034, B\"{o}blingen, Germany},
and M.~Mehring}

\affiliation{2. Physikalisches Institut, Universit\"at Stuttgart,
70550 Stuttgart, Germany}

\author{A.~Dul\v{c}i\'{c}}
\affiliation{Department of Physics, Faculty of Science,
University of Zagreb, POB 331, 10002 Zagreb, Croatia}
\date{September 6, 2002; revised version December 4, 2002}

\thanks{present address: D.-N. Peligrad: Philips Research
Laboratories, Weisshausstrasse 2, D-52066, Aachen, Germany; B.
Nebendahl: Agilent Technologies Deutschland GmbH, Herrenberger
Str. 130, D-71034, B\"{o}blingen, Germany}

\begin{abstract}

Critical fluctuations above $T_c$ are studied in the microwave complex
conductivity of BSCCO-2212, BSCCO-2223, and YBCO thin films. The analysis
of the experimental data yields the absolute values and the temperature
dependence of the reduced coherence length $\xi(T)/\xi_0$. Besides the well
known {\it 3D} XY critical regime having the static critical exponent
$\nu = 2/3$, a crossover to a new critical regime with $\nu = 1$ is observed
when $T_c$ is approached. In more anisotropic superconductors the reduced
coherence lengths are larger, and the critical regimes extend to higher
temperatures.\\
\end{abstract}


\maketitle

\par

The nature of superconducting fluctuations in high-temperature
superconductors has continued to attract a great deal of
attention. The small coherence lengths, high transition
temperatures, and large anisotropy due to the layered structure
make the fluctuations of the order parameter much stronger than in
classical low temperature superconductors. Particularly intriguing
has appeared the possibility of observing critical fluctuations in
a fairly wide temperature range around $T_c$. In this regime the
critical exponents deviate with respect to their mean-field
values. Renormalization group theory provides an appropriate
description of this phenomenon. One obtains scaling laws and
universality features which are of great importance in identifying
the nature of the phase transition \cite{Fisher:91}. The critical
fluctuations in high-$T_c$ superconductors have been studied by
penetration depth \cite{Kamal:94,Anlage:96}, specific heat
\cite{Salamon:93,Overend:94}, magnetization \cite{Salamon:93},
resistivity \cite{Menegotto:97,Han:98,Han:00}, thermal expansivity
\cite{Pasler:98,Meingast:01}, and two-coil inductive measurements
\cite{Osborn:02}. The conclusions drawn from these data seemed to point
at the critical behavior pertaining to the {\it 3D} XY universality
class in high-$T_c$ superconductors. However, the discrepancies in the
region of critical behavior varied from a fraction of a
degree \cite{Han:98,Han:00} to $\pm 10$ K \cite{Kamal:94,Pasler:98}.
\par
In this paper we present novel results and a new type of data
analysis of the microwave fluctuation conductivity in thin films of
high temperature superconductors $Bi_2Sr_2CaCu_2O_{8+\delta}$
(BSCCO-2212), $Bi_2Sr_2Ca_2Cu_3O_{10+\delta}$ (BSCCO-2223), and
$YBa_2Cu_3O_{7-\delta}$  (YBCO).
The advantage of the $\it{ac}$ method is that one obtains two
experimental data sets, i.~e. the real and imaginary parts of the
fluctuation conductivity, both of which have to corroborate with a
given theoretical model using the same set of parameters. It
represents therefore a more stringent test on the nature of the
fluctuations than methods yielding a single experimental
curve. Using this feature, we determine for the first time
experimentally the absolute values and the temperature dependence
of the coherence length above  $T_c$ in high-Tc superconductors.
Quite surprisingly, we observe in all our samples
two critical regimes with the static critical exponents $\nu =
2/3$ well above $T_c$, and a crossover to $\nu = 1$ when $T_c$ is
approached.

\par
YBCO thin films having 200~nm thickness were grown on $MgO$
substrate. BSCCO-2212 (350~nm) and BSCCO-2223 (100~nm) thin films
were grown on $LaAlO_3$ and $NdGaO_3$ substrates, respectively.
The sample was positioned in the center of an elliptical copper
cavity resonating in $_{\rm e}$TE$_{111}$ mode at $\approx
9.5$~GHz, with the microwave electric field $E_{\omega}$ parallel
to the {\it ab}-plane of the superconducting film. The temperature
of the sample could be varied from 2~K to room temperature by a
heater and sensor assembly mounted on the sapphire sample holder.
The $Q$-factor was measured by a recently introduced modulation
technique \cite{Nebendahl:01}, which enables the resolution of
$\Delta (1/2Q)$ to 0.02~ppm, while the frequency shift was
monitored by a microwave frequency counter.
\begin{figure}
\centerline{\includegraphics[width=0.7\textwidth]{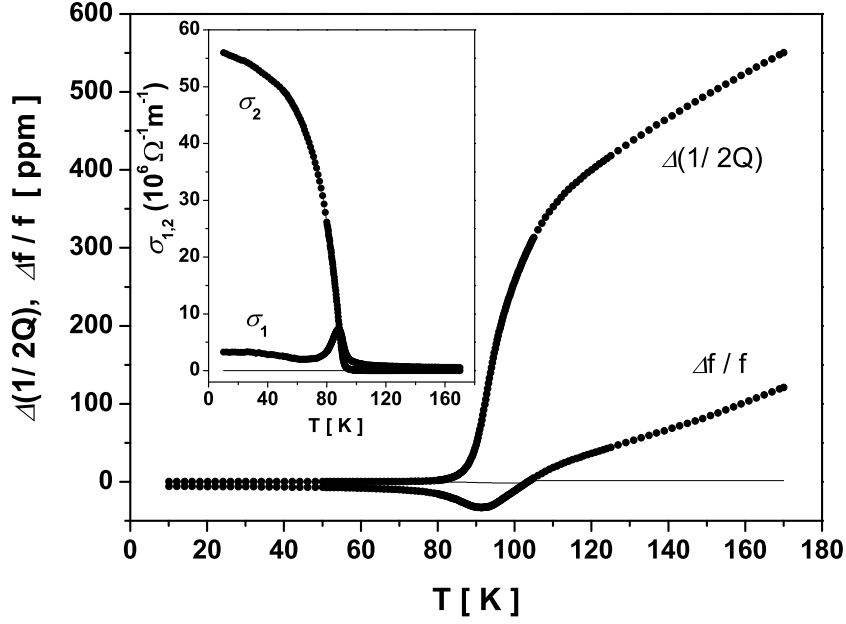}}
\caption{Temperature dependences of the experimentally measured
$\Delta(1/2Q)$ and $\Delta f/f$ in a BSCCO-2212 thin film. The
inset shows the real ($\sigma_1$) and imaginary ($\sigma_2$) parts
of the complex conductivity.} \label{Fig1}
\end{figure}
Fig.~\ref{Fig1} shows the experimental values of $\Delta (1/2Q)$
and $\Delta f/f$ for the BSCCO-2212 thin film. The complex
frequency shift $\Delta \widetilde{\omega}/\omega = \Delta f/f +
i\,\Delta (1/2Q)$ is related to the sample and cavity parameters
according to the cavity perturbation analysis \cite{Peligrad:01}
\begin{equation}
\displaystyle \frac{\Delta \widetilde{\omega}}{\omega}=
\frac{\Gamma}{N}\left[1-N+\frac{(\widetilde{k}/k_0)^2 N}
{\left[\coth(i\widetilde{k}d/2)+\tanh(i\widetilde{k}\zeta)\right]
i\widetilde{k}d/2}\right]^{-1}\, ,
\label{1}
\end{equation}
where $\Gamma$ is the filling factor of the sample in the cavity,
and $N$ is the depolarization factor of the film. The complex
wavevector in the film is $\displaystyle \widetilde{k} = k_0
\sqrt{1 -i \displaystyle
{\widetilde{\sigma}/(\epsilon_0\omega)}}$, where $\displaystyle
k_0=\omega\sqrt{\mu_0\epsilon_0}$ is the vacuum wavevector, and
$\widetilde{\sigma} = \sigma_1 - i \sigma_2$ is the complex
conductivity of the film. The thickness of the film is $d$, and
$\zeta$ is the asymmetry parameter due to the substrate
\cite{Peligrad:01}. The parameters $N$ and $\Gamma$ in
Eq.~(\ref{1}) have been evaluated from the comparison of the
experimental and theoretical curves of $\Delta (1/2Q)$ and $\Delta
f/f$ in the normal state far above $T_c$ where $\sigma_n$ is known
from $\it{dc}$ measurements. By taking the ratio of the slopes of
the theoretical $\Delta (1/2Q)$ and $\Delta f/f$ curves, one
eliminates $\Gamma$ so that $N$ can be determined from the
comparison to the experimental ratio. In the next step, $\Gamma$
can be determined from the measured value of $\Delta (1/2Q)$ at a
given temperature in the normal state. We have used $\sigma_n =
6.4~\cdot 10^5~\Omega^{-1} m^{-1}$ for the normal state
conductivity at 150~K. Similar evaluation at temperatures above
150~K did not change the resulting parameters. Eq.~(\ref{1}) can
then be used to convert the experimental data for $\Delta (1/2Q)$
and $\Delta f/f$ at any temperature to obtain the corresponding
experimental values of $\sigma_1$ and $\sigma_2$ as shown in the
inset to Fig.~\ref{Fig1}. Similar results for the complex
conductivity in BSCCO-2212 have been obtained previously on single
crystals \cite{Waldram:99}. The observed peak in $\sigma_1$ is due
to the superconducting fluctuations and its maximum is reached
when the coherence length diverges \cite{Fisher:91,Dorsey:91}. We
found $T_c = 87.9$~K in our BSCCO-2212 thin film.
\par
\begin{figure}
\centerline{\includegraphics[width=0.7\textwidth]{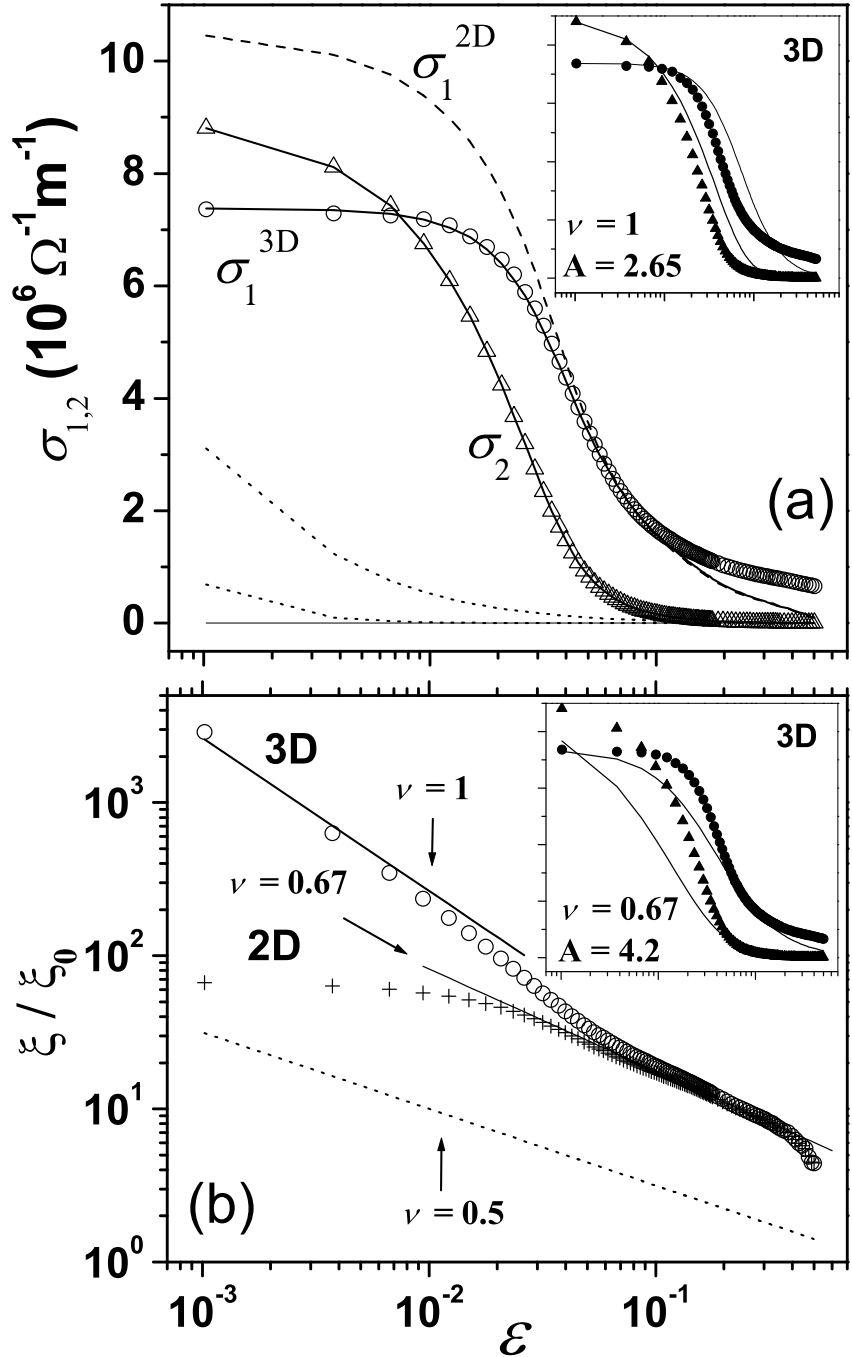}}
\caption{(a) Temperature dependences of the total measured real
($\bigcirc$) and imaginary ($\bigtriangleup$) part of the complex
conductivity above $T_c$ in a BSCCO-2212 thin film. The dotted
lines are the conductivity calculated using the Gaussian form of
the coherence length. The full and dashed lines are the results of
the selfconsistent {\it 3D} and {\it 2D} calculations,
respectively, as explained in the text. (b) The reduced coherence
length deduced from the experimental data of $\sigma_2$ by means
of the {\it 3D} ($\bigcirc$) and {\it 2D} ($+$) theoretical
expressions. The full lines show the slopes $\nu = 1$ and $\nu =
2/3$ for the two critical regimes. The Gaussian form
$1/\sqrt{\epsilon}$ is shown by the dotted lines. The insets show
the experimental data compared with the {\it 3D} calculations
using the coherence lengths of the form $(\xi(T)/\xi_0) =
A/\epsilon^{\nu}$.} \label{Fig2}
\end{figure}

In this paper we focus on $\sigma_1$ and $\sigma_2$ at, and above
$T_c$, shown on an expanded scale in Fig.~\ref{Fig2}(a). Here the
reduced temperature $\epsilon = \ln\left({T/T_c}\right)$ is used.
The $\it{ac}$ fluctuation conductivity above $T_c$ is given by
\cite{Peligrad:02}
\begin{equation}
\displaystyle {\widetilde {\sigma}}^{\hspace{0.05cm}{\it 3D}} =
\frac{e^2}{32\hbar\xi_{0c}}\left(\frac{\xi(T)}{\xi_0}\right)
\int_0^{Q_{ab}}\int_{-Q_c}^{Q_c}
\frac{4\, q_{ab}^{\,3}\, \left[1 - i \Omega (1+q_{ab}^2+q_c^2)^{-1}\right]}
{\pi (1+q_{ab}^2+q_c^2)[\Omega^2+(1+q_{ab}^2+q_c^2)^2]}
\,d q_{ab}\,d q_c \,\,\, .
\label{3}
\end{equation}
The prefactor is the Aslamazov-Larkin term for the {\it 3D} case. The reduced
coherence length $\xi(T)/\xi_0$ is the same for the in-plane and c-axis
coherence lengths, i.~e. $(\xi_{ab}(T)/\xi_{0ab}) = (\xi_c(T)/\xi_{0c})$,
so that we use it without subscripts. The cutoff in the fluctuation wavevector
is introduced in $\displaystyle Q_{ab}(T) = k_{ab}^{max} \xi_{ab}(T) =
\sqrt{2} \Lambda_{ab}\,\xi_{ab}(T)/ \xi_{0ab}$ for the ab-plane and
$\displaystyle Q_c(T) = k_c^{max} \xi_c(T) = \Lambda_c\,\xi_c(T)/ \xi_{0c}$
along the c-axis, rather than a single cutoff on the modulus \cite{Silva:02}.
The dimensionless parameter $\displaystyle \Omega =
\left(\pi\, \hbar\,\omega/16\,k_B\,T_c\right)\left(\xi(T)/\xi_0\right)^2$
involves the operating microwave frequency $\omega$ and the temperature
dependence of the reduced coherence length. The analytical result of
the integration in Eq.~(\ref{3}) and its application in the data analysis
has been reported in detail \cite{Peligrad:02}. The cutoff parameters
$\Lambda_{ab}$ and $\Lambda_c$ are constrained by the ratio
$\sigma_2(T_c)/\sigma_1(T_c)$ which can be evaluated from the experimental
data. In the present case we found $\Lambda_{ab}=0.7$ and $\Lambda_c=0.06$.
Also, the parameter $\xi_{0c}$ in Eq.~(\ref{3}) can be determined from the
experimental value of $\sigma_2(T_c)$, and we obtained $\xi_{0c}=0.05$~nm for
the BSCCO-2212 sample. Similar results for $\xi_{0c}$ were obtained also
from $\it{dc}$ conductivity measurements \cite{Hopfengartner:91}.
\par
The temperature dependence in the $\it ac$ fluctuation
conductivity of Eq.~(\ref{3}) is due to the reduced coherence
length. It is interesting to examine first the simple Gaussian
form $(\xi(T)/\xi_0) = 1/\sqrt{\epsilon}$. The fluctuation
conductivity calculated using the Gaussian coherence length is
shown by the dotted lines in Fig.~\ref{Fig2}(a). Since the
experimental values are much higher one may conclude that the
fluctuations are not Gaussian in this temperature range. We have
tried also other forms of the type $(\xi(T)/\xi_0) =
A/\epsilon^{\nu}$, but none of those could fit satisfactorily the
experimental data. Some examples are shown in the insets of
Fig.~\ref{Fig2}. With the variation of the parameters $A$ and
$\nu$, one can fit better either the section close to $T_c$, or
the one at higher temperatures, or make a compromise which never
gets the right shape in the whole temperature region. Obviously, a
single static critical exponent $\nu$ cannot describe the
experimental results. The true behavior of $\xi(T)/\xi_0$ has to
be determined from the experimental data itself. Therefore, we
have used the experimental values of $\sigma_2(T)$ to solve
numerically for $\xi(T)/\xi_0$ in the imaginary part of
$\displaystyle {\widetilde {\sigma}}^{\hspace{0.05cm}{\it 3D}}$
given by Eq.~(\ref{3}). The results are shown in
Fig.~\ref{Fig2}(b). One observes that in the region $0.07 <
\epsilon < 0.35$ the static critical exponent is close to $\nu =
2/3$ as expected for the critical regime of the {\it 3D} XY
universality class. Thus, the present analysis recovers, in this
part, the results obtained previously by other techniques in
high-$T_c$ superconductors. The solution of Eq.~(\ref{3}) it seems
to yield correct results when used in the critical regime. Namely,
the recent calculation by Wickham et al. \cite{Wickham:00} has
shown that the form of the expression for the $\it ac$ fluctuation
conductivity does not depend on the regime of the fluctuations.
Only the coherence length has different behavior in the Gaussian
and critical regimes. In the present case, the full result in
Fig.~\ref{Fig2}(b) shows a clear crossover from the {\it 3D} XY
critical regime to another critical regime with $\nu = 1$ when
$T_c$ is approached. The appearance of these two critical regimes
with a well defined crossover has not been reported previously.
The absolute values of the reduced coherence lenght reveal that
the crossover between the regimes is not just a change of the
critical exponent but also involves a step in the magnitude of the
coherence length.
\par
On the high temperature side, one observes in Fig.~\ref{Fig2}(b) a
deviation from the {\it 3D} XY critical regime at $\epsilon > 0.35$,
probably marking the crossover to the Gaussian regime with the critical
exponent $\nu = 1/2$. The drop seen in the last few points at
$\epsilon > 0.5$ can be interpreted in terms of the energy cutoff in the
fluctuation spectrum \cite{Carballeira:01,Vina:02}. This high temperature
region is not in the focus of the present report.
\par
The extraction of the coherence length was based on the imaginary
part of the $\it{ac}$ fluctuation conductivity $\sigma_2$ which is
due only to the superconducting fluctuations, i.~e. it has no
contribution from the normal electrons. We may now check the
consistency of the analysis on the real part of  the $\it{ac}$
fluctuation conductivity. By inserting the values of
$\xi(T)/\xi_0$ from Fig.~\ref{Fig2}(b) in the real part of the
solution of Eq.~(\ref{3}), we can calculate the corresponding
values of $\sigma_1$. The result is plotted as the full line in
Fig.~\ref{Fig2}(a). One observes an excellent agreement between
the calculated and the total measured real conductivity up to
$\epsilon = 0.09$. Beyond this region, the calculated values of
$\sigma_1$ fall below the total experimental $\sigma_1^{tot}$. It
appears that for $\epsilon < 0.09$, the normal conductivity
contribution vanishes due to the decrease of the one-electron
density of states at the Fermi level \cite{Timusk:99,Varlamov:99}.
However, for $\epsilon > 0.09$, there appears gradually a
contribution of the normal conductivity which has to be added to
the fluctuation conductivity $\sigma_1$ in order to get
$\sigma_1^{tot}$. A more detailed analysis of this effect will be
reported elsewhere \cite{Peligrad:02c}.
\par

We have also analyzed the same experimental data of the BSCCO-2212
sample using the expression for the $\it{ac}$ fluctuation
conductivity in the  {\it 2D} case \cite{Peligrad:02,Silva:02}.
Following the procedure analogous to the {\it 3D} case described
above, we find the values of $\xi(T)/\xi_0$ for the {\it 2D} case
which are also shown in Fig.~\ref{Fig2}(b). One finds that this coherence
length saturates when $T_c$ is approached. Since this is a completely
unphysical behavior, one may conclude that BSCCO-2212 is not an effective
{\it 2D} system near $T_c$. However, the analysis at higher temperatures
reveals that both, {\it 2D} and {\it 3D} calculations yield nearly the same slope of 2/3.
The effective layer thickness in the {\it 2D } expression for the fluctuation
conductivity is a free parameter which affects strongly
the absolute values but only slightly the slopes of $\xi(T)/\xi_0$. In Fig.~\ref{Fig2}(b)
we have used the effective layer thickness of 2.8~nm, which yields the best fit
to the $\sigma_1$ data at higher temperatures, and in addition matches the coherence
 length of the {\it 3D} case. This remarkable
result shows that the anisotropic {\it 3D} behavior matches with
the {\it 2D} behavior when the coherence length $\xi_c(T)$
approaches the cutoff value $\xi_{0c}/\Lambda_c$.
\par
One can check also the calculated $\sigma_1$ for the {\it 2D} case. The
dashed line in Fig.~\ref{Fig2}(a) shows a strong deviation from the
experimental data at temperatures close to $T_c$. This proves again that
the behavior in this temperature range is not of the {\it 2D} nature. At
higher temperatures, however, the calculated $\sigma_1$ in the {\it 2D} and
{\it 3D} cases practically overlap, thus corroborating the result found
in the coherence length.
\par
\begin{figure}
\centerline{\includegraphics[width=0.7\textwidth]{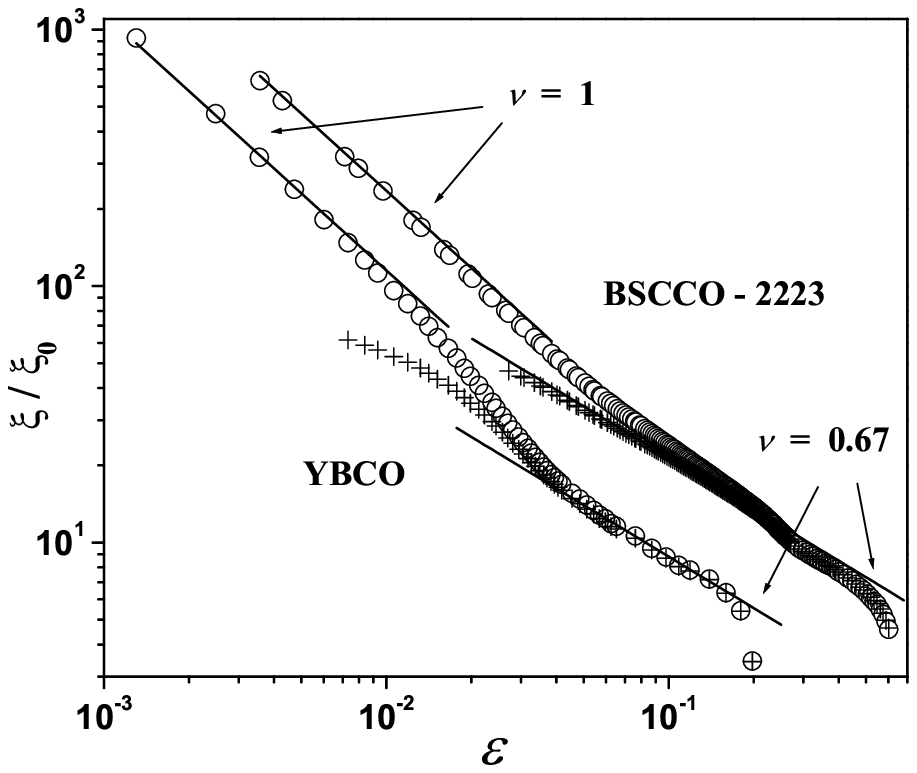}}
\caption{Reduced coherence lengths in the high-$T_c$
superconductors YBCO and BSCCO-2223 with various degrees of
anisotropy. Both, 3D($\bigcirc$) and 2D($+$) results are
presented.} \label{Fig3}
\end{figure}
Fig.~\ref{Fig3} shows the final results of the analysis of the
fluctuation conductivity  in YBCO, and BSCCO-2223 films. From the
temperature dependences of the reduced coherence lengths one
concludes readily that the appearance of multiple critical regimes
of the superconducting fluctuations is a general feature in
high-$T_c$ layered superconductors. The anisotropy of the system
affects the temperature ranges of the critical regimes. The more
anisotropic superconductors have their critical regimes extended
to higher temperatures. BSCCO-2223 exhibits the critical state
with $\nu = 1$ up to 1.4~K above $T_c$, which is considerably more
extended than found above in BSCCO-2212. The {\it 3D} XY critical
regime with $\nu = 2/3$ is found in BSCCO-2223 in a large interval
$0.08 < \epsilon < 0.44$.
\par
YBCO is the least anisotropic material of the three
superconductors studied here. We observe in Fig.~\ref{Fig3} that
it also develops the critical regime with $\nu = 1$, but very
close to $T_c$. The values of the reduced coherence length
$\xi(T)/\xi_0$ are much lower than in the BSCCO superconductors.
Also, the $\it 3D$ XY critical regime with $\nu = 2/3$ is found in
YBCO in a much narrower temperature range $0.05 < \epsilon <
0.12$. This finding is in agreement with the combined analysis of
the $\it{dc}$ conductivity, specific heat, and susceptibility
measurements in YBCO single crystals \cite{Ramallo:99}. Beyond
$\epsilon = 0.12$, the reduced coherence length in YBCO shows a
drop towards the Gaussian value $1/\sqrt{\epsilon}$. With such low
values of $\xi(T)/\xi_0$ the $\it{ac}$ fluctuation conductivity
becomes so small that our signal to noise ratio precludes a
further analysis.
\par
In conclusion, we have determined for the first time the absolute values and
the temperature dependence of the reduced coherence length $\xi(T)/\xi_0$
above $T_c$ in high-$T_c$ superconductors. Our results show in detail how
the fluctuations of the order parameter develop in YBCO, BSCCO-2212,
and BSCCO-2223 systems with different degrees of anisotropy. In
the vicinity of $T_c$, one finds a hitherto unobserved critical regime with
the static critical exponent $\nu = 1$. At higher temperatures one observes
a crossover to a critical regime having $\nu = 2/3$ as predicted for the {\it 3D} XY
universality class.
\par
The authors acknowledge the preparation of the BSCCO-2223 thin film to
Dr. A. Attenberger, and V. R\v{a}dulescu for the help in the
development of the required software for data analysis.


\end{document}